\documentclass[12pt]{article}
\usepackage{graphicx,fancyhdr}

\begin{document}
%\begin{abstract}
%\end{abstract}
%\tableofcontents
\def\K{\mathord{\cal K}}
\def\la{\langle}
\def\ra{\rangle}
\def\ltsim{\mathop{\,<\kern-1.05em\lower1.ex\hbox{$\sim$}\,}}
\def\gtsim{\mathop{\,>\kern-1.05em\lower1.ex\hbox{$\sim$}\,}}

\begin{center}

%Draft: 1 (20.11.2003)

\subsection*{Single-ion versus exchange anisotropy in calculating anisotropic
susceptibilities of thin ferromagnetic Heisenberg films within many-body
Green's function theory}
\vspace{2cm}

P. Fr\"obrich$^+$, and P.J. Kuntz
\vspace{1cm}

Hahn-Meitner-Institut Berlin, Glienicker Stra{\ss}e 100, D-14109 Berlin,
Germany,\\
$^+$also: Institut f\"ur Theoretische Physik, Freie Universit\"at Berlin\\
Arnimallee 14, D-14195 Berlin, Germany\\
\end{center}
\vspace{3cm}

{\bf Abstract.}
We compare transverse and parallel static susceptibilities of in-plane uniaxial
anisotropic ferromagnetic Heisenberg films calculated in the framework of
many-body Green's function theory using
single-ion anisotropies with the previously investigated case of exchange
anisotropies. On the basis of the
calculated observables (easy and hard axes magnetizations and susceptibilities)
no significant differences are found, i.e.
it is {\em not} possible to propose an experiment that might decide which kind
of anisotropy is acting in an actual ferromagnetic film.

%The many-body Green's function theory developed in our previous work for
%treating the reorientation of the magnetization of thin ferromagnetic films
%is extended to include the exchange anisotropy. This leads to additional
%momentum dependencies which require some non-trivial changes in the formalism.
%The theory
%is developed for arbitrary spin values S and for multilayers.
%The effects of the exchange anisotropy and the single-ion
%anisotropy, which was treated in our earlier work, on the magnetic
%properties of thin ferromagnetic films are compared.
\vspace{2cm}

{\bf PACS.} 75.10.Jm Quantized spin models - 75.30.Ds Spin waves - 75.70.Ak
Magnetic properties of monolayers and thin films

\newpage
\subsection*{1. Introduction}

Jensen et al. \cite{JKWO03} have measured parallel and tranverse
susceptibilities of a bi-layer Co film with an {\em in-plane} uniaxial
anisotropy, and analysed their results
with a many-body Green's function theory assuming a spin value of $S=1/2$.
We have generalized their work to multilayers and arbitrary spin in Ref.
\cite{FK03}. In both papers an exchange anisotropy was used, because it is
easier to treat than the single-ion anisotropy. In connection with the
reorientation of the magnetization of a ferromagnetic film (with an
{\em out-of-plane} anisotropy) as function of the temperature and film
thickness
we have already discussed similarities and differences between single-ion and
exchange anisotropies \cite{FKEPJB03}. In the latter paper the magnetic
dipole-dipole interaction was also included.

In the present paper we calculate within the Green's function
formalism anisotropic in-plane susceptibilities using the single-ion
anisotropy, and compare with the results of Ref. \cite{FK03}, where the
exchange anisotropy was used. Although we have shown in Ref. \cite{FKS02}
how the single-ion anisotropy can be treated exactly (for any strength of the
anisotropy) by introducing higher-order Green's functions, the application to
multilayers and $S>1$ is very
cumbersome. This is not the case when using, as we did in Refs. \cite{FKEPJB03}
and \cite{FJKE00} and we do in the present paper, an approximate decoupling on
the level of the lowest-order Green's functions proposed by Anderson and Callen
\cite{AC64}, which however is only a good approximation for small anisotropies,
as we showed in Ref. \cite{HFKTJ02} by comparing with `exact' Quantum Monte
Carlo calculations. In keeping with Refs. \cite{JKWO03}
and \cite{FK03} we do not include the dipole-dipole interaction, because it is
nearly isotropic for an in-plane situation.

The paper is organized as follows. In Section 2 we explain the model
and establish the Green's function formalism. Section 3 displays the numerical
results. In Section 4 we summarize the results and present our conclusions.

\subsection*{2. The model and the Green's function formalism}

Although the general formalism is rather similar to our previous work we
repeat it here to make the paper self-contained.

The Hamiltonian we use in this paper
consists of an isotropic Heisenberg exchange
interaction with strength $J_{kl}$ between nearest neighbour lattice sites,
 a second-order {\em in-plane} single-ion lattice
anisotropy with strength $K_{2,k}$,  and an external magnetic
field ${\bf B}=(B^x,B^y,B^z)$:
\begin{eqnarray}
{\cal
H}=&-&\frac{1}{2}\sum_{<kl>}J_{kl}(S_k^-S_l^++S_k^zS_l^z)
-\sum_kK_{2,k}(S_k^z)^2\nonumber\\
&-&\sum_k\Big(\frac{1}{2}B^-S_k^++\frac{1}{2}B^+S_k^-+B^zS_k^z\Big).
\label{5.1}
\end{eqnarray}
Here the notation $S_k^{\pm}=S_k^x\pm iS_k^y$ and $B^{\pm}=B^x\pm iB^y$ is
introduced, where $k$ and $l$ are lattice site indices and $<kl>$ indicates
summation over nearest neighbours only. The in-plane lattice directions are the
x and z-axes. The field $B^y$ will be put to zero lateron.

In order to treat the  problem for general spin $S$, we need the
following Green's functions
\begin{equation}
G_{ij,\eta}^{\alpha,mn}(\omega)=\la\la
S_i^\alpha;(S_j^z)^m(S_j^-)^n\ra\ra_{\omega,\eta}\ ,
\label{5.2}
\end{equation}
where $\alpha=(+,-,z)$ takes care of all directions in space, $\eta=\pm 1$
refers to the anticommutator or commutator Green's functions, respectively, and
$n\geq 1, m\geq 0$ are positive integers, necessary for dealing with higher
spin values $S$.

The exact equations of motion are
\begin{equation}
\omega G_{ij,\eta}^{\alpha,mn}(\omega)=A_{ij,\eta}^{\alpha,mn}+\la\la
[S_i^\alpha,{\cal H}]_-;(S_j^z)^m(S_j^-)^n\ra\ra_{\omega,\eta}
\label{5.3}
\end{equation}
with the inhomogeneities
\begin{equation}
A_{ij,\eta}^{\alpha,mn}=\la[S_i^\alpha,(S_j^z)^m(S_j^-)^n]_{\eta}\ra,
\label{5.4}
\end{equation}
where $\la ...\ra=Tr(...e^{-\beta{\cal H}})/Tr(e^{-\beta{\cal H}})$. The equations are given
explicitly by %
\begin{eqnarray}
\omega G_{ij,\eta}^{\pm,mn}&=&A_{ij,\eta}^{\pm,mn}\nonumber\\
& &\mp\sum_{k}J_{ik}\Big(\la\la S_i^zS_k^\pm;(S_j^z)^m(S_j^-)^n\ra\ra
-\la\la S_k^zS_i^\pm;(S_j^z)^m(S_j^-)^n\ra\ra\Big)\nonumber\\
& &\pm K_{2,i}\la\la(S_i^\pm
S_i^z+S_i^zS_i^\pm);(S_j^z)^m(S_j^-)^n\ra\ra\nonumber\\
& &\mp B^\pm G_{ij,\eta}^{z,mn}\pm B^zG_{ij,\eta}^{\pm,mn}\nonumber\\
\omega G_{ij,\eta}^{z,mn}&=&A_{ij(\eta)}^{z,mn}\nonumber\\
& &+\frac{1}{2}\sum_kJ_{ik}\la\la(S_i^-S_k^+-S_k^-S_i^+);
(S_j^z)^m(S_j^-)^n\ra\ra\nonumber\\
& &-\frac{1}{2}B^- G_{ij,\eta}^{+,mn}+\frac{1}{2}B^+G_{ij,\eta}^{-,mn}.
\label{5.5}
\end{eqnarray}

After solving these equations the components of the magnetization can be
determined from the Green's functions via the spectral theorem.
A solution is possible by establishing a closed system of equations by
decoupling the higher-order Green's functions on the right-hand sides.
Contrary to Ref. \cite{FKS02}, where we proceed to higher-order Green's
functions, we stay here at the level of the lowest-order equations.
For the exchange-interaction  terms, we use a
generalized Tyablikov- (or RPA-) decoupling
\begin{equation}
\la\la S_i^\alpha S_k^\beta;(S_j^z)^m(S_j^-)^n\ra\ra_\eta \simeq\la
S_i^\alpha\ra
G_{kj,\eta}^{\beta,mn}+\la S_k^\beta\ra G_{ij,\eta}^{\alpha,mn} .
\label{5.6}
\end{equation}
The terms from the single-ion anisotropy have to be decoupled differently,
because an RPA decoupling leads to unphysical results; e.g. for spin $S=1/2$,
the terms due to the single-ion anisotropy do not vanish in RPA, as they should
do, because in this case $\sum_i K_{2,i}\la (S_i^z)^2\ra$ is a constant and
should not influence the equations of motion.
In the appendix of Ref. \cite{FJK00} we investigated different decoupling
schemes proposed in the literature, e.g. those of Lines \cite{Lin67} or that of
Anderson and Callen \cite{AC64}, which should be reasonable for single-ion
anisotropies small compared to the exchange interaction. We found the
Anderson-Callen decoupling to be
most adequate. It consists in implementing the suggestion of Callen
\cite{Cal63} to improve the RPA by treating the diagonal terms arising from
the single-ion anisotropy as well. This leads to
\begin{eqnarray}
& &\la\la(S_i^\pm S_i^z+S_i^zS_i^\pm);(S_j^z)^m(S_j^-)^n\ra\ra_\eta \nonumber\\
& &\simeq 2\la S_i^z\ra\Big(1-\frac{1}{2S^2}[S(S+1)-\la
S_i^zS_i^z\ra]\Big)G_{ij,\eta}^{\pm,mn}.
\label{5.7}
\end{eqnarray}
This term vanishes for $S=1/2$ as it should.

After a Fourier transform to momentum space, one obtains, for a
ferromagnetic film with $N$ layers,
$3N$ equations of motion for a $3N$-dimensional Green's function vector ${\bf
G}^{mn}$:
\begin{equation}
(\omega{\bf 1}-{\bf \Gamma}){\bf G}^{mn}={\bf A}^{mn},
\label{5.8}
\end{equation}
where ${\bf 1}$ is the
$3N\times 3N$ unit matrix. The Green's function vectors and inhomogeneity
vectors each
consist of $N$  three-dimensional subvectors which are characterized by the
layer indices $i$ and $j$

\begin{equation}
{\bf G}_{ij}^{mn}({\bf{k},\omega})\  =
\left( \begin{array}{c}
G_{ij}^{+,mn}({\bf{k}},\omega) \\ G_{ij}^{-,mn}({\bf{k}},\omega)  \\
G_{ij}^{z,mn}({\bf{k}},\omega)
\end{array} \right), \hspace{0.5cm}
{\bf A}_{ij}^{mn} {=}
 \left( \begin{array}{c} A_{ij}^{+,mn} \\ A_{ij}^{-,mn} \\
A_{ij}^{z,mn} \end{array} \right) \;.
\label{5.9} \end{equation}

The equations of motion are then expressed in terms of these layer vectors, and
$3\times 3 $ submatrices ${\bf \Gamma}_{ij}$ of the $3N\times 3N$
matrix ${\bf\Gamma}$
\begin{equation}
\left[ \omega {\bf 1}-\left( \begin{array}{cccc}
{\bf\Gamma}_{11} & {\bf\Gamma}_{12} & \ldots & {\bf\Gamma}_{1N} \\
{\bf\Gamma}_{21} & {\bf\Gamma}_{22} & \ldots & {\bf\Gamma}_{2N} \\
\ldots & \ldots & \ldots & \ldots \\
{\bf\Gamma}_{N1} & {\bf\Gamma}_{N2} & \ldots & {\bf\Gamma}_{NN}
\end{array}\right)\right]\left[ \begin{array}{c}
{\bf G}_{1j} \\ {\bf G}_{2j} \\ \ldots \\ {\bf G}_{Nj} \end{array}
\right]=\left[ \begin{array}{c}
{\bf A}_{1j}\delta_{1j} \\ {\bf A}_{2j}\delta_{2j} \\ \ldots \\
{\bf A}_{Nj}\delta_{Nj} \end{array}
\right] \;, \hspace{0.5cm} j=1,...,N\;.
\label{5.10}
\end{equation}
After applying the decoupling procedures  (\ref{5.6}) and (\ref{5.7}),
the  ${\bf \Gamma}$ matrix reduces to a band matrix with zeros in the
${\bf \Gamma}_{ij}$ sub-matrices, when $j>i+1$ and $j<i-1$.
The  diagonal sub-matrices ${\bf \Gamma}_{ii}$ are of size $3\times 3$
and have the form
\begin{equation}
 {\bf \Gamma}_{ii}= \left( \begin{array}
{@{\hspace*{3mm}}c@{\hspace*{5mm}}c@{\hspace*{5mm}}c@{\hspace*{3mm}}}
\;\;\;H^z_i & 0 & -H^+_i \\ 0 & -H^z_i & \;\;\;H^-_i \\
-\frac{1}{2}{H}^-_i & \;\frac{1}{2}{H}^+_i & 0
\end{array} \right)
\ . \label{5.11}
\end{equation}
where
\begin{eqnarray}
H^z_i&=&Z_i+\la S_i^z\ra J_{ii}(q-\gamma_{\bf k})\ ,
\nonumber\\
Z_i&=&B^z_i
+J_{i,i+1}\la S_{i+1}^{z}\ra+J_{i,i-1}\la
S_{i-1}^{z}\ra\nonumber\\
& &+K_{2,i}2\la S_i^z\ra
\Big(1-\frac{1}{2S^2}[S(S+1)-\la S_i^zS_i^z\ra]\Big)\ ,
\nonumber \\
{H}^\pm_i&=&B^\pm_i+\la S_i^\pm\ra J_{ii}(q-\gamma_{\bf
k})
+J_{i,i+1}\la S_{i+1}^{\pm}\ra+J_{i,i-1}\la
S_{i-1}^{\pm}\ra \ .
\label{5.12}
\end{eqnarray}
For a square lattice, to which we restrict ourselves in the present paper, and
a lattice constant
taken to be unity, $\gamma_{\bf k}=2(\cos k_x+\cos k_y)$, and $q=4$ is the
number of intra-layer nearest neighbours.

The $3\times 3$
off-diagonal sub-matrices ${\bf \Gamma}_{ij}$ for $j= i\pm 1$ are of the
form %
\begin{equation}
 {\bf \Gamma}_{ij} = \left( \begin{array}
{@{\hspace*{3mm}}c@{\hspace*{5mm}}c@{\hspace*{5mm}}c@{\hspace*{3mm}}}
-J_{ij}\la S_i^z\ra & 0 & \;\;\;J_{ij}\la S_i^+\ra \\
0 & \;\;J_{ij}\la S_i^z\ra & -J_{ij}\la S_i^-\ra \\
\frac{1}{2}J_{ij}\la S_i^-\ra &
-\frac{1}{2}J_{ij}\la S_i^+\ra & 0 \end{array} \right) \;.
\label{5.13}
\end{equation}

When treating the monolayer, one can use the spectral theorem for calculating
the components of the magnetization. This was done in Ref.
\cite{FJK00} for the case of spin $S=1$ and an out-of-plane single-ion
anisotropy by using the commutator Green's functions.
In order to obtain sufficient equations it was necessary, to add equations
coming
from the condition that the commutator Green's functions have to be regular at
$\omega=0$, which we call the regularity conditions.

The treatment of multilayers is only practicable when
using the eigenvector method
developed in Ref. \cite{FJKE00}.
The essential features are as follows.
One starts with a transformation, which diagonalizes the ${\bf \Gamma}$-matrix
of equation (\ref{5.8})
\begin{equation}
{\bf L\Gamma R}={\bf \Omega},
\label{5.14}
\end{equation}
where ${\bf \Omega}$ is a diagonal matrix with eigenvalues $\omega_{\tau}$
($\tau=1,..., 3N$). For the problem above it turns out that there is one
eigenvalue equal to zero for each layer, which has to be
handled appropriately. The transformation matrix {\bf R} and
its inverse ${\bf R}^{-1}={\bf L}$ are obtained from the right eigenvectors of
${\bf \Gamma}$
as columns and from the left eigenvectors as rows, respectively. These matrices
are normalized to unity: {\bf RL}={\bf LR}={\bf 1}.

Multiplying the equation of motion (\ref{5.8}) from the left by {\bf L} and
inserting {\bf 1}={\bf RL} one finds
\begin{equation}
(\omega{\bf 1}-{\bf \Omega}){\bf L}{\bf G}_\eta^{mn}={\bf LA}_\eta^{mn}.
\label{5.15}
\end{equation}
Defining ${\cal G}_\eta^{mn}={\bf LG}_\eta^{mn}$
and ${\cal A}_\eta^{mn}={\bf LA}_\eta^{mn}$ one obtains
\begin{equation}
(\omega {\bf 1}-{\bf \Omega}){\cal G}_\eta^{mn}={\cal A}_\eta^{mn}.
\label{5.16}
\end{equation}
${\cal G}_\eta^{mn}$ is a vector of new Green's functions, each component
$\tau$ of which has but a single pole
\begin{equation}
{\cal G}_\eta^{mn,\tau}=\frac{{\cal A}_\eta^{mn,\tau}}{\omega-\omega_\tau}\ .
\label{5.17}
\end{equation}
This is the important point because it allows application of the spectral
theorem, e.g. \cite{GHE01}, to
each component separately. We obtain for the component $\tau$ of correlation
vector ${\cal C}^{mn}={\bf L}{\bf C}^{mn}$
( where ${\bf C}^{mn}=\la (S^z)^m(S^-)^nS^\alpha\ra$ with $(\alpha=+,-,z)$)
\begin{equation}
{\cal C}^{mn,\tau}=\frac{{\cal
A}_{\eta}^{mn,\tau}}{e^{\beta\omega_\tau}+\eta}+\frac{1}{2}(1-\eta)\frac{1}{2}
\lim_{\omega\rightarrow 0}\omega \frac{{\cal
A}_{\eta=+1}^{mn,\tau}}{\omega-\omega_\tau} .
\label{5.18}
\end{equation}
We emphasize that when ($\eta=-1$), the second term of this
equation, which is due to the anticommutator Green's function, has to be taken
into account. This term occurs for $\omega_\tau=0$ and can be simplified by
using the relation between anticommutator and commutator
\begin{equation}
{\cal A}_{\eta=+1}^{mn,0}={\cal A}^{mn,0}_{\eta=-1}+2{\cal
C}^{mn,0}={\bf L}_0(A_{\eta=-1}^{mn}+2{\bf C}^{mn}),
\label{5.19}
\end{equation}
where the index $\tau=0$ refers to the eigenvector with $\omega_\tau=0$.

The term ${\bf L}_0A_{\eta=-1}^{mn}=0$ vanishes due to the fact that the
commutator Green's function has to be regular at the origin
\begin{equation}
\lim_{\omega\rightarrow 0}\omega G_{\eta=-1}^{\alpha,mn}=0,
\label{5.20}
\end{equation}
which leads to the regularity conditions:
\begin{equation}
{H}^xA_{\eta=-1}^{+,mn}+{H}^xA_{\eta=-1}^{-,mn}+2H^zA_{\eta=-1}^{z,
m n } =0.
\label{5.21}
\end{equation}
For details, see Ref. \cite{FJKE00}.

This is equivalent to
\begin{equation}
{\bf L}_0A_{\eta=-1}^{mn}=0,
\label{5.22}
\end{equation}
because the left eigenvector of the ${\bf \Gamma}$-matrix with eigenvector zero
has the structure
\begin{equation}
{\bf L}_0\propto ({H}^x, {H}^x, 2H^z),
\label{5.23}
\end{equation}
what can be seen analytically.
For more details concerning the use of the regularity conditions, see Refs.
\cite{FK03,FJKE00}.

We mention an alternative method, published in Ref. \cite{FKPRB03}, of treating
zero
eigenvalues occurring in the equation of motion matrix, which is based on a
singular value decomposition of this matrix, and where there is no need
for the use of the anticommutator Green's function.

The equations for the correlations are obtained by multiplying equation
(\ref{5.18})  from the left with ${\bf R}$ and using equation (\ref{5.22});
i.e.
\begin{equation}
{\bf C}={\bf R}{\bf {\cal E}}{\bf L}{\bf A}+{\bf R}_0{\bf L}_0{\bf C},
\label{5.24}
\end{equation}
where ${\bf {\cal E}}$ is a diagonal matrix with matrix elements
${\cal E}_{ij}=\delta_{ij}(e^{\beta\omega_i}-1)^{-1}$ for eigenvalues
$\omega_i\neq 0$, and $0$ for eigenvalues $\omega_i=0$.

This set of equations has to be solved self-consistently together with the
regularity conditions
(\ref{5.21}). This determines the magnetizations and the moments of the
magnetizations $\la (S^z)^n\ra$ with $n=2S+1$ for the highest moment, $S$
being the spin quantum number.
For details see Appendix A of Ref. \cite{FKEPJB03}, where an analogous set of
similar equations is given more explicitly for the case of the out-of-plane
situation.

The susceptibilities with respect to the easy ($\chi_{zz}$) and hard
($\chi_{xx}$) axes are calculated as differential quotients
\begin{eqnarray}
\chi_{zz}&=&\Big(\la S^z(B^z)\ra -\la S^z(0)\ra\Big)/B^z\nonumber\\
\chi_{xx}&=&\Big(\la S^x(B^x)\ra -\la S^x(0)\ra\Big)/B^x,
\label{5.25}
\end{eqnarray}
where the use of $B^{z(x)}=0.01/S$ turns out to be small enough, see also Ref.
\cite{FK03}.

\subsection*{3. Numerical results}

In this section we show numerical results obtained with the
single-ion
anisotropy in comparison with that from the exchange anisotropy, for which the
relevant equations were derived in Ref. \cite{FK03} .
As the single-ion anisotropy is not active for $S=1/2$ we will show
results for $S\geq 1$. In an attempt to obtain universal
(independent of the spin quantum number S) curves, we have scaled the
parameters ($J,D,B^{x(z)}$) entering the Hamiltonian
of Ref. \cite{FK03} as $\tilde{J}/S(S+1)=J$, $\tilde{D}/S(S+1)=D$
($D$ being the strength of the exchange anisotropy),
$\tilde{B}^{x(z)}/S=B^{x(z)}$. In the present paper we use an additional
scaling for the strength of the single-ion anisotropy
$\tilde{K_2}/(S-1/2)=K_2$. This has been proven to be the proper scaling in
Ref.\cite{FJKE00}, because it leads to the correct limit
$\lim_{T\rightarrow 0}(K_2(T)/K_2(0))=1$, when calculating the temperature-
dependent anisotropy by minimizing the free energy with respect to the
equilibrium orientation angle of the magnetization.

\subsubsection*{3.1 The monolayer with arbitrary spin}
In order to compare results obtained with the single-ion anisotropy and with
the exchange anisotropy we fit the strength of the single-ion anisotropy to
$K_2=5.625$ for a square lattice spin S=1 monolayer such that the easy axis
magnetization
$\la S^z\ra/S$ comes as close as possible to the magnetization obtained
previously \cite{FK03} with the exchange anisotropy chosen to be $D=5$. The
exchange interaction parameter is $J=100$, and a small magnetic field in
x-direction is used, $B^x=0.01/S$, which stabilizes the calculation. The
comparison is shown in Fig.1.

\begin{figure}[htb]
\begin{center}
\protect
\includegraphics*[bb = 80  90 510 410,
angle=0,clip=true,width=11cm]{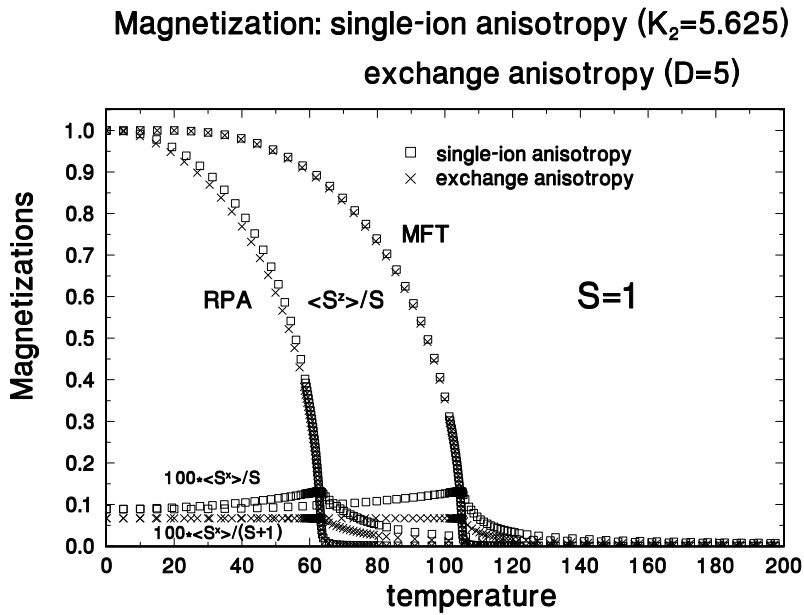}
\protect
\caption{The magnetization $\la S^z\ra/S$  of a
ferromagnetic spin $S=1$ Heisenberg monolayer for a square lattice is shown as
function of the temperature. Comparison
is made between Green's function (RPA) calculations using the exchange
anisotropy ($D=5$, crosses) and the single-ion anisotropy ($K_2=5.625$,
open squares) with Anderson-Callen decoupling. The corresponding results of
mean field (MFT)  calculations are also displayed. Also shown are the
quantities
$100*\la S^x\ra/(S+1)$ for the exchange anisotropy and $100*\la S^x\ra/S$ for
the single-ion anisotropy; the factor 100 is introduced to make the curves
visible.} \label{fig1}
\end{center}
\end{figure}
It is surprising that the results for the easy axis magnetization
$\la S^z\ra$ are very
similar over the whole temperature range although the physical origin for the
anisotropies is very different. An analogous result was observed for the out-of
plane situation discussed in Ref. \cite{FKEPJB03}. The agreement is not so good
for
the hard axis magnetization, which is a constant for the exchange
anisotropy for temperatures below the Curie temperature, whereas it rises
slightly up to the Curie temperature when using the single-ion anisotropy.
In Ref.\cite{FK03} it was shown analytically that the hard axis magnetization
is universal for a scaling $\la S^x\ra/(S+1)$ when using the exchange
anisotropy. For the single-ion anisotropy a scaling $\la S^x\ra/S$ is found to
be more appropriate.
Comparison is made
also with the corresponding mean field (MFT) calculations, obtained by putting
$\gamma_{\bf k}=0$ in eqn (\ref{5.12}), showing the well known shift
to larger Curie temperatures (by a factor of about two for the monolayer) due
to the missing magnon correlations.

In Figs. 2 and 3 we show the easy and hard axes magnetizations for a monolayer
with
different spin values $S$. Whereas one observes in Fig.2 a nearly perfect
scaling for
the RPA calculations  with the exchange anisotropy ($S=1/2,1,3/2,2,3,4,6,13/2$,
from Ref.\cite{FK03} ) and a universal Curie temperature $T_C(S)$  for RPA and
MFT, this is not the case for the corresponding results with
the single-ion anisotropy shown for $S=1,3/2,4,5$ in Fig.3, although the
violation of scaling is not dramatical.
\begin{figure}[htb]
\begin{center}
\protect
\includegraphics*[bb = 80  85 510 410,
angle=0,clip=true,width=10cm]{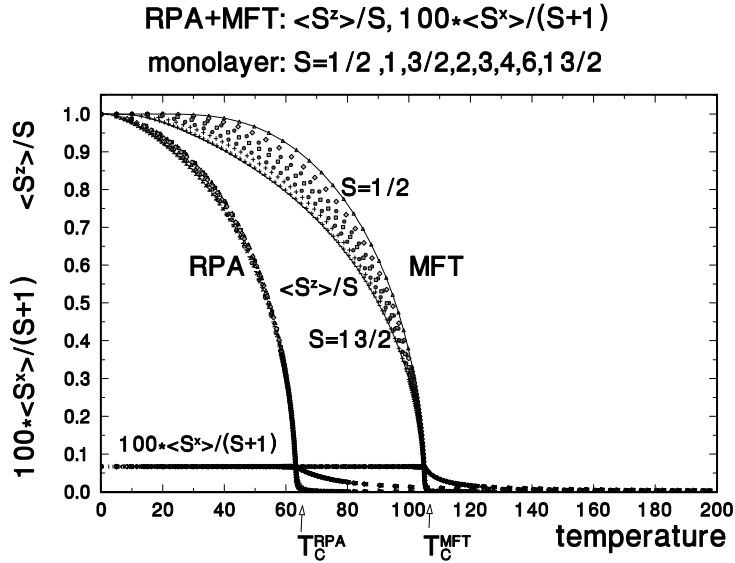}
\protect
\caption{The magnetizations $\la S^z\ra/S$  of
spin $S=1/2,1,3/2,2,3,4,6,13/2$ Heisenberg monolayers for a square lattice are
shown as functions of the temperature, from Ref. \cite{FK03}. Comparison
is made between Green's function (RPA) calculations  and  results of
mean field theory (MFT), using the exchange
anisotropy strength, $D=5$. Also shown is the hard axis magnetization,
which scales to a universal curve when using $100*\la S^x\ra/(S+1)$,
where the factor 100 is introduced to make the curves visible.}
\label{fig2}
\end{center}
\end{figure}
\newpage
.
\begin{figure}[htb]
\begin{center}
\protect
\includegraphics*[bb = 80  85 510 410,
angle=0,clip=true,width=10cm]{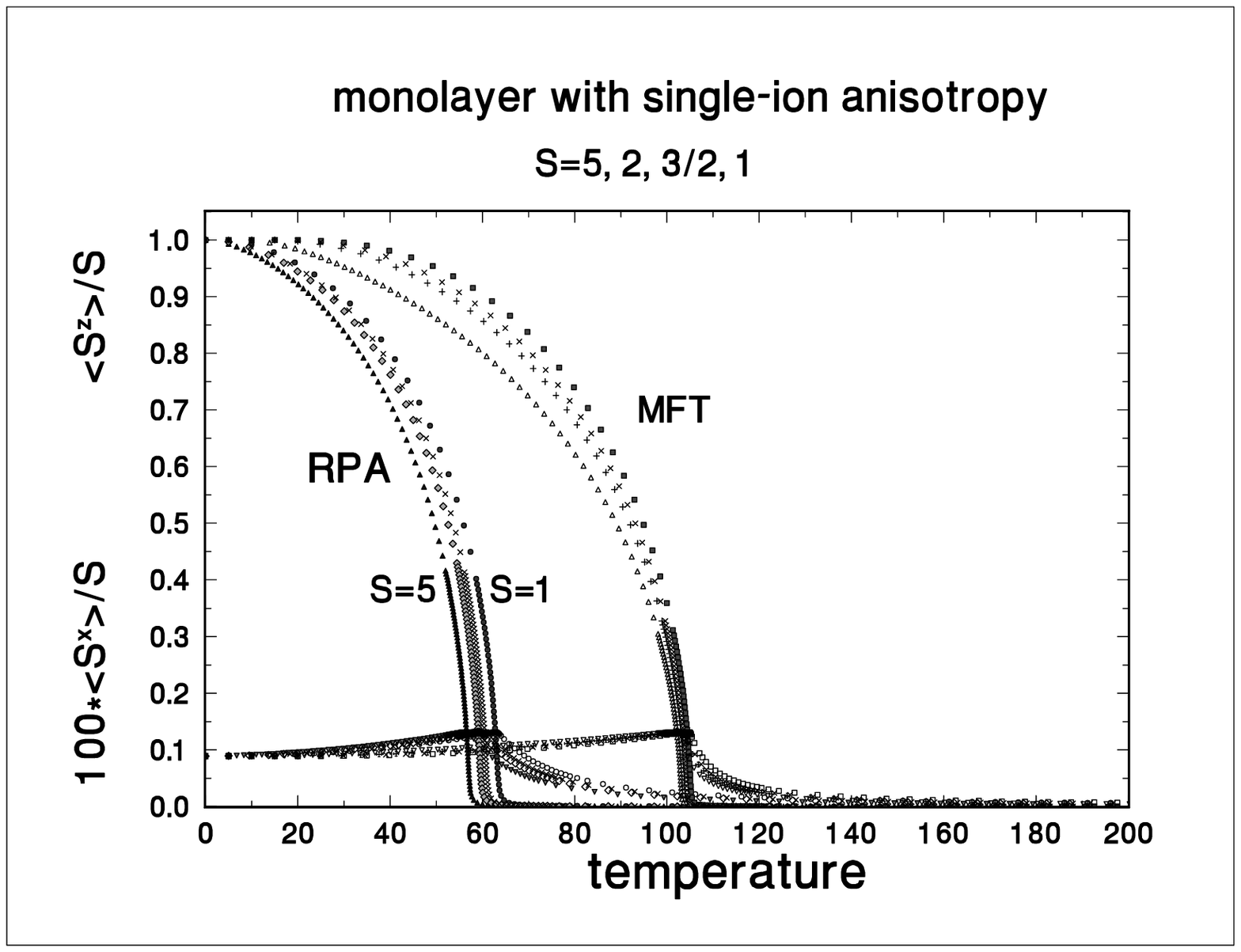}
\protect
\caption{The magnetizations $\la S^z\ra/S$  of
ferromagnetic spin $S=1,2,3/2,5$ Heisenberg monolayers for a square lattice are
shown as functions of the temperature. Comparison
is made between Green's function (RPA) calculations using  the single-ion
anisotropy strength of $K_2=5.625$,
and the corresponding results of
mean field theory (MFT). Also shown are the quantities
 $100*\la S^x\ra/S$;
the factor 100 is introduced to make the curves visible. There is only an
approximate scaling behaviour.}
\label{fig3}
\end{center}
\end{figure}

Turning to the inverse easy and hard axes susceptibilities $\chi_{zz}^{-1}$ and
$\chi_{xx}^{-1}$ we find very similar results for the exchange anisotropy and
the single-ion anisotropy. In particular in the paramagnetic region
($T>T_{\rm Curie}$) one has a curved behaviour as function of the temperature
for the
susceptibilities in RPA due to the presence of spin waves, whereas the
corresponding
MFT calculations show a Curie-Weiss (linear in the temperature) behaviour due
to missing magnon excitations. One observes a slightly less universal behaviour
for the results for the single-ion anisotropy, in Fig. 4 and 5, when comparing
with the results of the exchange anisotropy, see Figs. 2 and 3 of Ref.
\cite{FK03}. This is connected with the fact that using the exchange anisotropy
one finds universal values for the Curie temperatures $T_C^{RPA}(S)$ and
$T_C^{MFT}(S)$, which is not strictly the case when using the single-ion
anisotropy, see Fig. \ref{fig3}.We were also able to show analytically in
Ref. \cite{FK03} that $\chi_{xx}^{-1}*S(S+1)$ is universal for $T<T_C$ when
using the exchange anisotropy; this is not the case for the
single-ion anisotropy.
The only difference concerns the curves for the not perfectly scaled
RPA results for $\chi_{zz}^{-1}$: with the exchange anisotropy the
curve with the lowest spin value is left from the curves with the higher spin
values, whereas the inverse is true for the exchange anisotropy, but this is
not a very pronounced effect, and does not lead to a significant
difference between the results for the various anisotropies.

\begin{figure}[htb]
\begin{center}
\protect
\includegraphics*[bb = 80  85 510 410,
angle=0,clip=true,width=10cm]{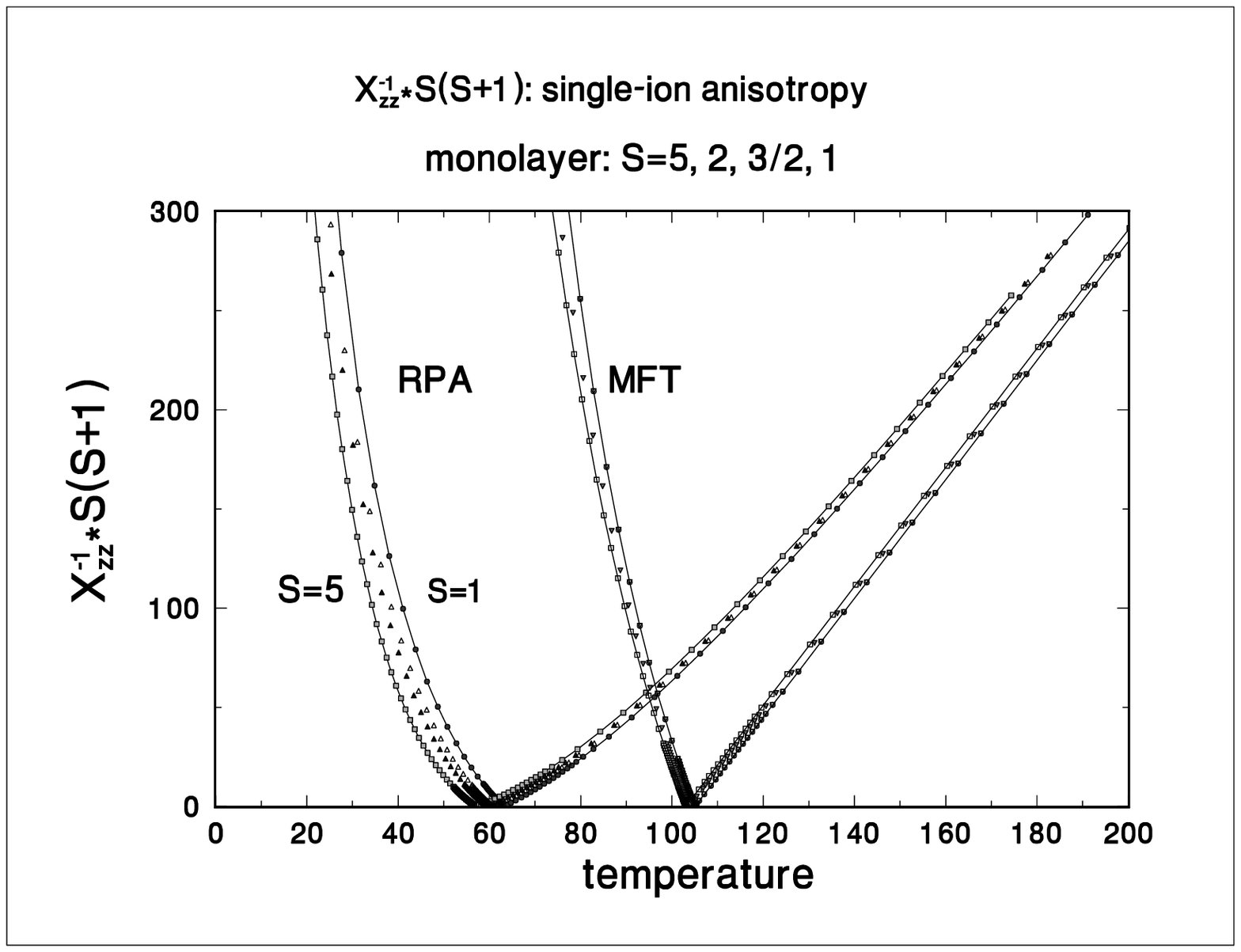}
\protect
\caption{`Universal' inverse easy axis susceptibilities
$\chi_{zz}^{-1}*S(S+1)$ of an in-plane anisotropic (due to the single-ion
anisotropy) ferromagnetic square lattice Heisenberg monolayer as functions of
the temperature for
spins $S=5, 2, 3/2, 1$. Comparison
is made between Green's function (RPA)  and
mean field (MFT)  calculations.}
\label{fig4}
\end{center}
\end{figure}
\newpage
.
\begin{figure}[htb]
\begin{center}
\protect
\includegraphics*[bb = 80  85 510 410,
angle=0,clip=true,width=10cm]{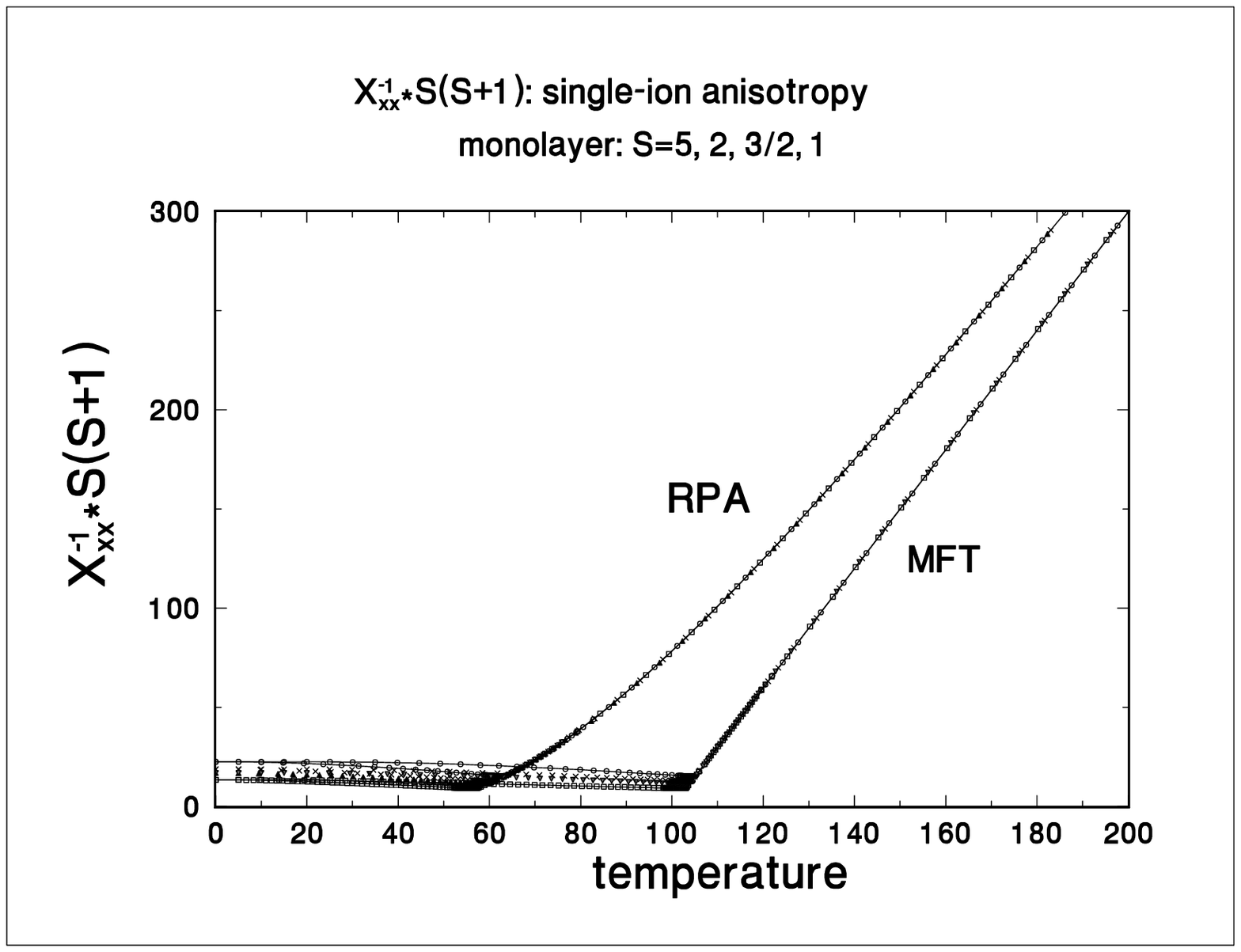}
\protect
\caption{`Universal' inverse hard axis susceptibilities
$\chi_{xx}^{-1}*S(S+1)$ of an in-plane anisotropic (due to the single-ion
anisotropy) ferromagnetic square lattice Heisenberg monolayer as functions of
the temperature for
spins $S=5, 2, 3/2, 1$. Comparison
is made between Green's function (RPA)  and
mean field (MFT)  calculations.
}
\label{fig5}
\end{center}
\end{figure}
\subsubsection*{3.2 Multilayers at fixed spin $S=1$}
In discussing multilayers with the exchange anisotropy we have considered only
the case of $S=1/2$ in Ref. \cite{FK03}.
In order to compare with results from the single-ion anisotropy we have to use
a larger spin value because $S=1/2$ is not active in this case.
We restrict ourselves to spin S=1 in the following.
We have performed also calculations with $S>1$ which scale with respect to the
spin in the same way as in the monolayer case.

In Fig. 6 we compare the Curie temperatures for $S=1$ multilayers for exchange
and single-ion anisotropies in RPA and MFT, using for each layer the same
parameters as for the monolayer. Remember that the parameters were fixed
such that the Curie temperatures for both anisotropies coincide for the
monolayer.
The Curie temperatures  for the multilayers $N=2, ..., 19$ (for N=19 one
is already close to the bulk limit) are only slightly lower for the single-ion
anisotropy than those for the exchange anisotropy.

\begin{figure}[htb]
\begin{center}
\protect
\includegraphics*[bb = 80  85 510 410,
angle=0,clip=true,width=10cm]{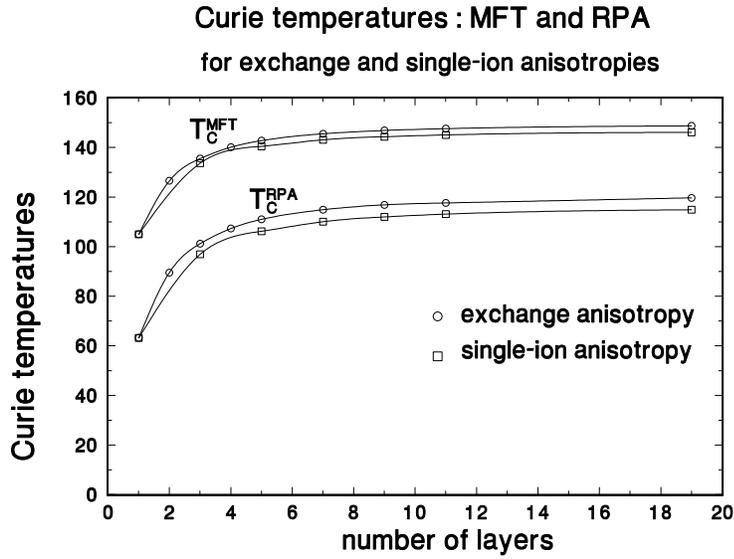}
\protect
\caption{Curie temperatures of ferromagnetic spin S=1 multilayers are shown as
function of the film thickness foe RPA and MFT using the exchange (open
circles) and the single-ion (black square) anisotropies.}
\label{fig6}
\end{center}
\end{figure}

In Figs. 7 and 8 we compare easy and hard axes inverse susceptibilities
calculated with single-ion and exchange anisotropy also for the multilayer
case.
In order to avoid cluttering the figures we restrict ourselves to a multilayer
with N=9 layers and spin $S=1$. For $N>9$ the corresponding curves would shift
only
slightly in accordance with the saturation of $T_C$, see Fig. \ref{fig6},
with
increasing film thickness. We display only the RPA results for the multilayer
(N=9) and compare with the RPA monolayer (N=1) result. Again there is no
significant difference in
the results for both anisotropies. We do not plot the corresponding mean field
results which
are shifted to higher temperatures and show in the paramagnetic region only a
linear in T Curie-Weiss behaviour, whereas the RPA results have curved
shapes due to the influence of magnon correlations, which are
completely absent in MFT.
\begin{figure}[htb]
\begin{center}
\protect
\includegraphics*[bb = 80  85 510 410,
angle=0,clip=true,width=10cm]{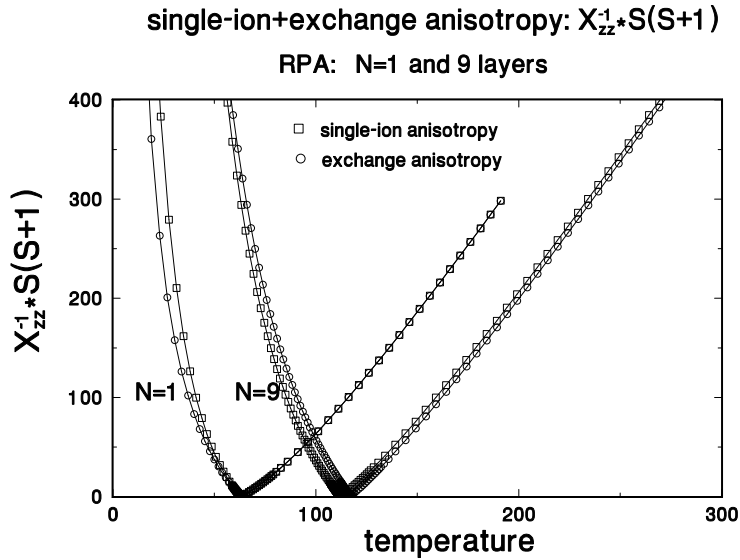}
\protect
\caption{The inverse easy axis susceptibilities $\chi_{zz}^{-1}$ of
ferromagnetic films in RPA
for spin $S=1$ for a monolayer (N=1) and a multilayer (N=19) as functions of
the temperature for single-ion and exchange anisotropies.}
\label{fig7}
\end{center}
\end{figure}
\begin{figure}[htb]
\begin{center}
\protect
\includegraphics*[bb = 80  85 510 410,
angle=0,clip=true,width=10cm]{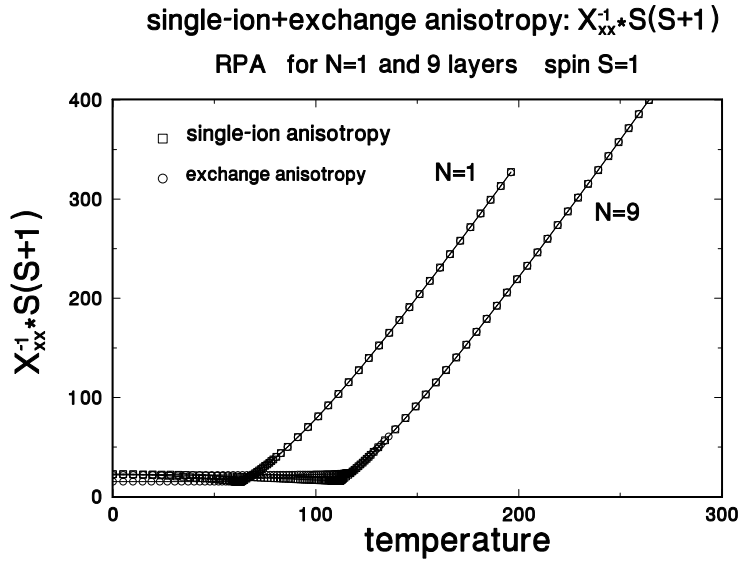}
\protect
\caption{The inverse hard axis susceptibilities $\chi_{xx}^{-1}$ of
ferromagnetic films in RPA
for spin $S=1$ for a monolayer (N=1) and a multilayer (N=19) as functions of
the temperature for single-ion and exchange anisotropies.}
\label{fig8}
\end{center}
\end{figure}
\newpage

\subsection*{4. Summary and conclusions}
We have applied in this paper a many-body Green's function formalim to
calculate in-plane anisotropic static susceptibilities of ferromagnetic
Heisenberg films using the single-ion anisotropy, and compared with previous
calculations \cite{FK03} in which an exchange anisotropy was used. Although
both
kinds of anisotropies are of very different physical origin, it is possible, by
fitting the strengths of the anisotropies properly, to obtain nearly identical
values for the easy axis magnetizations over the complete temperature range for
an $S=1$ monolayer. Using the parameters obtained in this way also for
monolayers with
higher spin values and for multilayers, we looked for differences in the
results of calculations
with both kinds of anisotropies.

By using scaled variables we find a fairly
universal (independent of the spin quantum number S)
of easy and hard axes magnetizations and inverse susceptibilities. Universality
is better established for the exchange anisotropy; e.g. we find a universal
Curie temperature $T_C(S)$ for RPA and MFT. The scaling is not as perfect for
the single-ion anisotropy, but there are {\em no} dramatic deviations, which
might lead to a distinction of the influence of both anisotropies.
The general statement made in Ref. \cite{FK03} that it is sufficient to do a
calculation for a particular S and then to apply scaling to obtain the results
for other spin values, remains valid to a large extent also for the use of the
single-ion
anisotropy. It remains also true that the measurement of the
hard axis susceptibility gives in principle the possibility to obtain together
with a measurement of the Curie temperature information on the parameters of
the model, the exchange interaction and the anisotropy strengths.
One should, however, keep in mind that the quantitative results of the
present calculations are due to the use of a square lattice. They could change
significantly
by using different lattice types and also by  layer-dependent exchange
interactions and anisotropies. Such calculations are possible, because the
numerical program is written in such a way that layer-dependent coupling
constants can be used.

As a general result we state that our investigations up to now have {\em not}
lead to any significant differences  for the calculated observables
(easy and hard axes magnetizations and susceptibilities) when using on one hand
the single-ion anisotropy and on the other hand the exchange anisotropy.
Therefore it is not possible on the basis of our results to propose an
experiment that could decide  which kind of anisotropy is acting in an actual
ferromagnetic film.

\newpage .


\begin{thebibliography}{99}
%1
\bibitem{JKWO03} P.J. Jensen, S. Knappmann, W. Wulfhekel, H.P. Oepen,
Phys. Rev. B {\bf 67}, 184417 (2003).
%2
\bibitem{FK03} P. Fr\"obrich, P.J. Kuntz, cond-mat/0306243, submitted to Phys.
Rev. B
%3
\bibitem{FKEPJB03} P. Fr\"obrich, P.J. Kuntz, Eur. Phys. J. B {\bf 32}, 445
(2003).
%4
\bibitem{FKS02} P. Fr\"obrich, P.J. Kuntz, M. Saber, Ann. Phys. (Leipzig)
{\bf 11}, 387 (2002).
%5
\bibitem{FJKE00} P. Fr\"obrich, P.J. Jensen, P.J. Kuntz, A. Ecker, Eur. Phys.
J. B {\bf18}, 579 (2000).
%6
\bibitem{AC64} F.B. Anderson, H.B. Callen,  Phys. Rev. {\bf
136}, A1068 (1964).
%7
\bibitem{HFKTJ02} P. Henelius, P. Fr\"obrich, P.J. Kuntz, C. Timm, P.J. Jensen,
Phys. Rev. B {\bf 66}, 094407 (2002).
%8
\bibitem{FJK00} P. Fr\"obrich, P.J. Jensen, P.J. Kuntz, Eur. Phys. J. B
{\bf13}, 477 (2000).
%9
\bibitem{Lin67} M.E. Lines, Phys. Rev. {\bf156}, 534 (1967).
%10
\bibitem{Cal63} H.B. Callen, Phys. Rev. {\bf 130}, 890 (1963).
%11
\bibitem{GHE01} W. Gasser, E. Heiner, and K. Elk, in 'Greensche Funktionen in
der Festk\"orper- und Vielteilchenphysik', Wiley-VHC, Berlin, 2001, Chapter
3.3.
%12
\bibitem{FKPRB03} P. Fr\"obrich, P.J. Kuntz, Phys. Rev. B {\bf 68}, 014410
(2003).
%
\end{thebibliography}
\end{document}